\documentclass[12pt,preprint]{aastex}
\begin{document}
\title{Rotation of Cosmic Voids and Void-Spin Statistics}
\author{Jounghun Lee and Daeseong Park}
\affil{Department of Physics and Astronomy, FPRD, Seoul National University, 
Seoul 151-747, Korea} 
\email{jounghun@astro.snu.ac.kr,pds2001@astro.snu.ac.kr}
\begin{abstract}
We present a theoretical study of void spins and their correlation 
properties. The concept of the spin angular momentum for an unbound
void is introduced to quantify the effect of the tidal field on the
distribution of matter that make up the void. Both the analytical and
numerical approaches are used for our study. 
Analytically, we adopt the linear tidal torque model to evaluate 
the void spin-spin and spin-density correlations, assuming that a void forms 
in the initial region where the inertia momentum and the tidal shear tensors 
are maximally uncorrelated with each other. Numerically, we use the Millennium 
run galaxy catalog to find voids and calculate their spin statistics. 
The numerical results turn out to be in excellent agreement with the analytic 
predictions, both of which consistently show that there are strong spatial 
alignments between the spin axes of neighbor voids and strong
anti-alignments between the void spin axes and the directions to the
nearest voids. We expect that our work will provide a deeper insight
into the origin and properties of voids and the large scale structure.
\end{abstract}


\section{INTRODUCTION}

Recent galaxy redshift surveys \citep[e.g., 2dFGRs,][]{col-etal01}
have allowed us to map the universe on the largest scale ever and
study systematically the detailed properties of the cosmic large-scale
structures. It is confirmed by these surveys that the galaxies in the
universe are not distributed evenly but develop pattern of network
connected by filaments and sheets which in turn are separated by
immense voids.  

Voids are defined observationally as large regions with very low galaxy 
number density and shown to occupy approximately $40\%$ of the cosmic volume 
\citep{hoy-vog04}. The striking vastness of the observed void volumes 
inspired a flurry of research on the origin and properties of cosmic voids 
\citep{ela-pir97,sch-etal01,pee01,mat-whi03,ben-etal03,she-van04,hoy-vog04,
roj-etal04,col-etal05,hoy-etal05,got-etal05,sha-etal06}. It is now
generally believed that voids are originated from the local minima of
the primordial density fluctuations and expand faster than the rest
of the universe. 

The very fact that voids are extremely underdense and expand faster 
has led many authors to assume somewhat naively that the shapes of voids 
should be spherical. In fact previous analytical and numerical works on 
voids are based on this assumption 
\citep[e.g.,][]{dub-etal93,van-van93,fri-pir01,she-van04}. 

Very recently, however, \citet{sha-etal06} investigated systematically 
the shapes of voids by applying the excursion set approach to the 
density field in N-body simulations and demonstrated that the intrinsic 
shapes of voids are far from spherical symmetry. They claimed that voids 
are most vulnerable to external shears due to their low-density, so that 
their shapes are disturbed to be nonspherical in spite of their faster 
expansion. 

Although \citet{sha-etal06} focused on the shapes and morphology of voids, 
we note here that if the tidal shear plays a significant role in the case of 
voids, then it should cause not only the shapes of voids to be nonspherical 
but also voids to acquire the spin angular momentum. Our goal here is to 
construct an analytic model for the spin angular momentum of voids originated 
from the tidal shear effect and test the analytic predictions against the 
data from the recent Millennium run simulation. 

The organization of this paper is as follows. In $\S 2$, the analytic model 
for the void spin angular momentum and the void spin correlation properties 
are presented. In $\S 3$, the Millennium run galaxy catalog is analyzed 
to determine void spin statistics, and the numerical results are compared 
with the analytic predictions. In $\S 4$, the achievements and caveats of 
our works are discussed, and a final conclusion is drawn. 

\section{THE ANALYTIC MODEL}

Let us consider a Lagrangian region in the linear smoothed density 
field. According to the linear tidal torque theory \citep{dor70,whi84}, 
the spin angular momentum of matter which make up this region, 
${\bf J}\equiv (J_{i})$, is proportional to the tensor product as 
\begin{equation}
\label{eqn:ang}
J_{i} \propto \epsilon_{ijk}T_{jl}I_{lk},
\end{equation}
where ${\bf I}\equiv (I_{ij})$ is the inertia tensor of the region, and 
${\bf T}\equiv (T_{ij})$ is the local tidal tensor at the center of the mass 
of the region defined as the second derivative of the perturbation potential 
: $T_{ij}\equiv \partial_{i}\partial_{j}\Phi$.

Strictly speaking, equation (\ref{eqn:ang}) gives a valid first order
approximation to the spin angular momentum only when the Lagrangian
regions correspond to bound halos.  However, to quantify the effect of
tidal field on voids,  we extend the validity of the linear tidal
torque theory to the case of underdense proto-void regions and define
the {\it spin angular momentum of a void} by equation (\ref{eqn:ang}). 

Since a void is an unbound system, its spin angular momentum has a
distinct meaning from that of a bound halo although defined in a
similar way.  Basically, the spin angular momentum of a void is
introduced to quantify how a proto-void region reacts to the effect of
the local tidal force. The local tidal field will generate non-radial
motion  of the material in the proto-void  region, leading the
distribution of matter in the void to deviate from spherical symmetry.
The magnitude of the void spin angular momentum quantifies the degree
of  this deviation, and its direction indicates the direction
along which the maximum deviation occurs. 

It was shown by N-body simulations that in the overdense proto-halo regions 
the principal axes of ${\bf I}$ are in fact strongly correlated with that of 
${\bf T}$ \citep{lee-pen00,por-etal02}. Very recently, \citet{lee06} has 
shown analytically that the degree of the misalignment between the principal 
axes of ${\bf I}$ and ${\bf T}$ decreases monotonically with the linear 
density of the region. In other words, the more underdense a region is, 
the more misaligned the principal axes of ${\bf I}$ and ${\bf T}$ are. 

Given these numerical and analytical findings, we propose a new 
hypothesis that a void forms from the initial underdense region where the 
principal axes of ${\bf I}$ and ${\bf T}$ are maximally misaligned with 
each other. This hypothesis combined with equation (\ref{eqn:ang}) leads us 
to expect that the alignment between the void spin axes and the principal 
axes of the local tidal tensor should be strongest for the proto-void regions.

\citet{lee-pen00} showed that the alignment between the spin axis of a 
Lagrangian region and the intermediate principal axes of the local tidal 
tensor can be best quantified by calculating the expectation value of the 
unit spin given the tidal tensor, and they derived an approximate formula 
for it. Their original formula is characterized by a correlation parameter, 
$c$, in the range of $[0,1]$, which was introduced to represent the degree 
of the alignment \footnote{In the original formula, \citet{lee-pen00} used 
a rescaled correlation parameter, $a$, which is nothing but $3/5$ times $c$.} 
between the spin axes and the principal axes of the tidal tensors. 
The case of maximum alignment corresponds to $c=1$ while no alignment case 
to $c=0$. For the case of halo spins, they found $c\approx 0.3$ empirically.
In accordance of our hypothesis, $c=1$ for void spins.  

Now, with the value of $c$ set at unity, the formula of \citet{lee-pen00} is 
rewritten for the case of void spins as 
\begin{equation}
\label{eqn:spincorr}
\langle \hat{J}_{i}\hat{J}_{j}\vert\hat{\bf T}\rangle = 
\left(\frac{1}{3} + \frac{1}{5}\right)\delta_{ij} - 
\frac{3}{5}\hat{T}_{ik}\hat{T}_{kj},
\end{equation}
where $\hat{\bf J}\equiv{\bf J}/\vert{\bf J}\vert$ and 
$\hat{\bf T}\equiv \tilde{\bf T}/\vert\tilde{\bf T}\vert$ with 
$\tilde{T}_{ij}\equiv T_{ij}-\textrm{Tr}({\bf T})\delta_{ij}/3$.
Equation (\ref{eqn:spincorr}) indicates that the spin axes of voids 
are spatially correlated due to the spatial correlation of the tidal 
field. The alignments between the spin axes of neighbor void spins can be 
quantified by the two-point spin-spin correlation function which was 
introduced by \citet{pen-etal00} for the case of dark halos: 
\begin{equation}
\label{eqn:eta}
\eta_{V}(r) = \langle\vert\hat{\bf J}({\bf x})\cdot
\hat{\bf J}({\bf x + r})\vert^{2}\rangle - \frac{1}{3}.
\end{equation}
Here, the constant $1/3$ represents the value of 
$\langle\vert\hat{\bf J}({\bf x})\cdot\hat{\bf J}({\bf x}+{\bf r})
\vert^{2}\rangle$ for the case of no alignment. 

For the case of a power-law power spectrum, the void spin-spin correlation 
function, $\eta_{V}(r)$, can be approximated \citep{pen-etal00} as 
\begin{equation}
\label{eqn:spincorrft}
\eta_{V}(r) \approx \frac{3}{50}\frac{\xi^{2}_{R_V}(r)}{\xi^{2}_{R_V}(0)}.
\end{equation}
Here $\xi_{R_V}$ is the correlation function of the linear density 
field smoothed on a typical Lagrangian scale of voids, defined as 
\begin{equation}
\label{eqn:autocor}
\xi_{R_V}(r) \equiv \int_{-\infty}^{\infty}\Delta^{2}(k)\frac{\sin kr}{kr}
W^{2}(kR_{V})d\ln k,
\end{equation}
where $\Delta^{2}(k)$ is the dimensionless power spectrum and $W(kR_{V})$ 
is the top-hat spherical filter of scale radius, $R_V$. Since voids and 
galaxy clusters can be thought of the counterparts which are supposed to 
form at the local minima and maxima of the initial density field, 
respectively, we assume the typical Lagrangian scale of voids are same 
as that of galaxy clusters, setting the value of $R_{V}$ at $8h^{-1}$Mpc. 
Since the concordance $\Lambda$CDM cosmology is well represented by a 
power-law power spectrum $\Delta^{2}(k)\propto k$ on the scale of 
$8h^{-1}$Mpc, one can regard equation (\ref{eqn:eta}) as a valid 
approximation.

It is worth mentioning here that the misalignment between the principal
axes of  ${\bf T}$ and ${\bf I}$ (i.e., the value of the correlation
parameter $c=1$) is characteristic for those underdense regions in the
initial density field that will eventually become voids. Although
the majority of the initial underdense regions will evolve into voids,
some underdense regions could merge into overdense region to collapse
rather than to become voids \citep{col-etal05}.

\citet{lee-pen01} showed both analytically and numerically that for the 
case of dark halos the separation vectors between the neighbor halos, 
${\bf r}\equiv (r_{i})$, are aligned with the major axes of the tidal tensors. 
To quantify the degree of this alignment, \citet{lee-pen01} proposed the 
following formula: 
\begin{equation}
\label{eqn:dencorr}
\langle \hat{r}_{i}\hat{r}_{j}\vert\hat{\bf T}\rangle = 
\frac{1}{3}\delta_{ij} - b\left(\frac{1}{3}\delta_{ij}-
\hat{T}_{ik}\hat{T}_{kj}\right), 
\end{equation}
where $\hat{\bf r}\equiv {\bf r}/\vert{\bf r}\vert$ and $b$ is a correlation 
parameter in the range of $[0,1]$. The negative sign in front of the parameter 
$b$ indicates that $\hat{\bf r}$ is aligned with the {\it major} principal 
axis of $\hat{\bf T}$ unlike $\hat{\bf J}$. The value of $b$ was determined 
empirically to be approximately $0.3$ for the case of halos \citep{lee-pen01}. 

Similarly, we expect that the separation vectors between the neighbor 
voids are aligned with the {\it minor} axes of the tidal tensors.
Then, equation (\ref{eqn:dencorr}) should be also true for the case of
voids.  We assume further that $b$ has the same value of $0.3$. 

Now that the spin axes of voids are aligned with the intermediate axes 
of the local tidal tensors (eq.[\ref{eqn:spincorr}]) while the separation 
vectors to the nearest voids are aligned with the minor axes of the local 
tidal tensors, one can expect that the spin axes of voids should be 
anti-aligned with the separation vectors to the nearest voids.

The expected spin-direction anti-alignment can be quantified by the following 
correlation function \citep{lee-pen01}: 
\begin{equation}
\label{eqn:omega}
\omega_{V}(r) = \langle\vert\hat{\bf J}\cdot\hat{\bf r}\vert^{2}\rangle - 1/3.
\end{equation}
It was found by \citet{lee-pen01} that for the case of small-scale galactic 
halos equation (\ref{eqn:dencorrft}) can be approximated as 
$RR^{\prime}/(R^{2}+R^{\prime 2})$ where $R$ is the typical Lagrangian size 
of a galactic halo and $R^{\prime}\equiv R+r$. This approximation was made 
under the assumption that the tidal field does not vary significantly 
within the Lagrangian distance $r$. For the cases of large-scale voids, 
however, this assumption may not be valid since within the void separation 
the tidal field is likely to vary significantly. Taking into account the 
spatial variance of the tidal field, we find that for the case of voids 
the spin-direction correlation function is better approximated as 
\begin{equation}
\label{eqn:dencorrft}
\omega_{V}(r) \approx 
-\frac{3}{100}\frac{\xi^{2}_{R_V}(r)}{\xi^{2}_{R_V}(0)}.
\end{equation}

\section{RESULTS FROM SIMULATIONS}

\subsection{\it Identifying Voids}

To test the analytic predictions made in $\S 2$, we use the Millennium
run galaxy catalog carried out by the Virgo Consortium \citep{spr-etal05}. 
The catalog consists of total $8964936$ galaxies at redshift $z=0$ in a 
periodic box of linear size $500h^{-1}$Mpc, with information on various 
galaxy properties such as position, velocity, mass, magnitude, and
star formation rate. 

Voids are identified in the catalog by the void-finding algorithm described 
in \citet[][hereafter HV02]{hoy-vog02} which was in fact first suggested by 
\citet{ela-pir97}. To apply the HV02 algorithm to the Millennium run 
catalog, we first determine the values of the key parameters, $l$ and $s_{c}$. 
The two parameters, $l$ and $s_{c}$, represent the distance within
which a field galaxy is required not to have any three neighbors and a 
threshold for the minimum linear size of a void, respectively. 
For the Millennium run catalog, the key parameter values we obtain are 
$l=2.44h^{-1}$Mpc and $s_{c}=6h^{-1}$Mpc, respectively. For the
detailed description of the procedures to identify voids from the
Millennium run catalog, see our companion paper 
(D. Park \& J. Lee 2006 in preparation). 

The total $24037$ voids are found from the catalog. The volume $\Gamma_{V}$ 
of each void is measured by means of the Monte-Carlo integration method 
described in HV02. The effective radius of each void is then determined 
as $R_{e}\equiv (3\Gamma_{V}/4\pi)^{1/3}$. The mean value of $R_{e}$ 
is obtained to be $\bar{R}_{e}=10.45h^{-1}$Mpc.  

\begin{figure}
  \begin{center}
   \plotone{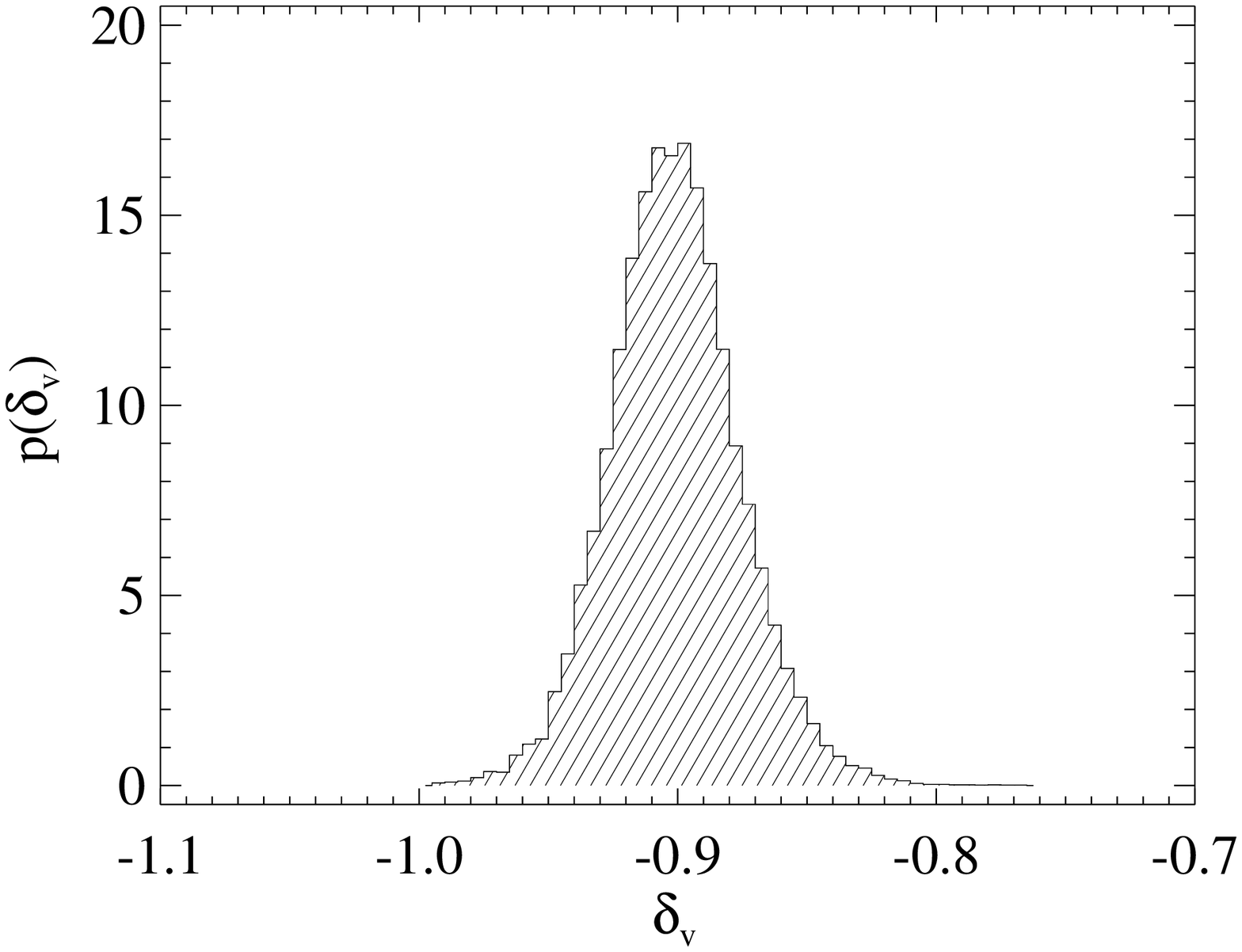}
\caption{Void density distribution measured from the Millennium run galaxy 
catalog.}
\label{fig:den}
 \end{center}
\end{figure}

The density of each void, $\delta_{V}$, is measured by counting the number 
of void galaxies: $\delta_{V} \equiv (n_{V}-\bar{n}_{g})/\bar{n}_{g}$ where 
$n_{V}\equiv N_{V}/\Gamma_{V}$ is the number density of void galaxies, and 
$\bar{n}_{g}$ is the mean number density of the total catalog galaxies. 
Figure \ref{fig:den} plots the probability distribution of $\delta_{V}$. 
As can be seen, the voids found in the catalog are extremely underdense 
with the mean density $\bar{\delta}_{V}=-0.92$. 

\subsection{\it Measuring the Void Spin Angular Momentum}

\begin{figure}
  \begin{center}
   \plotone{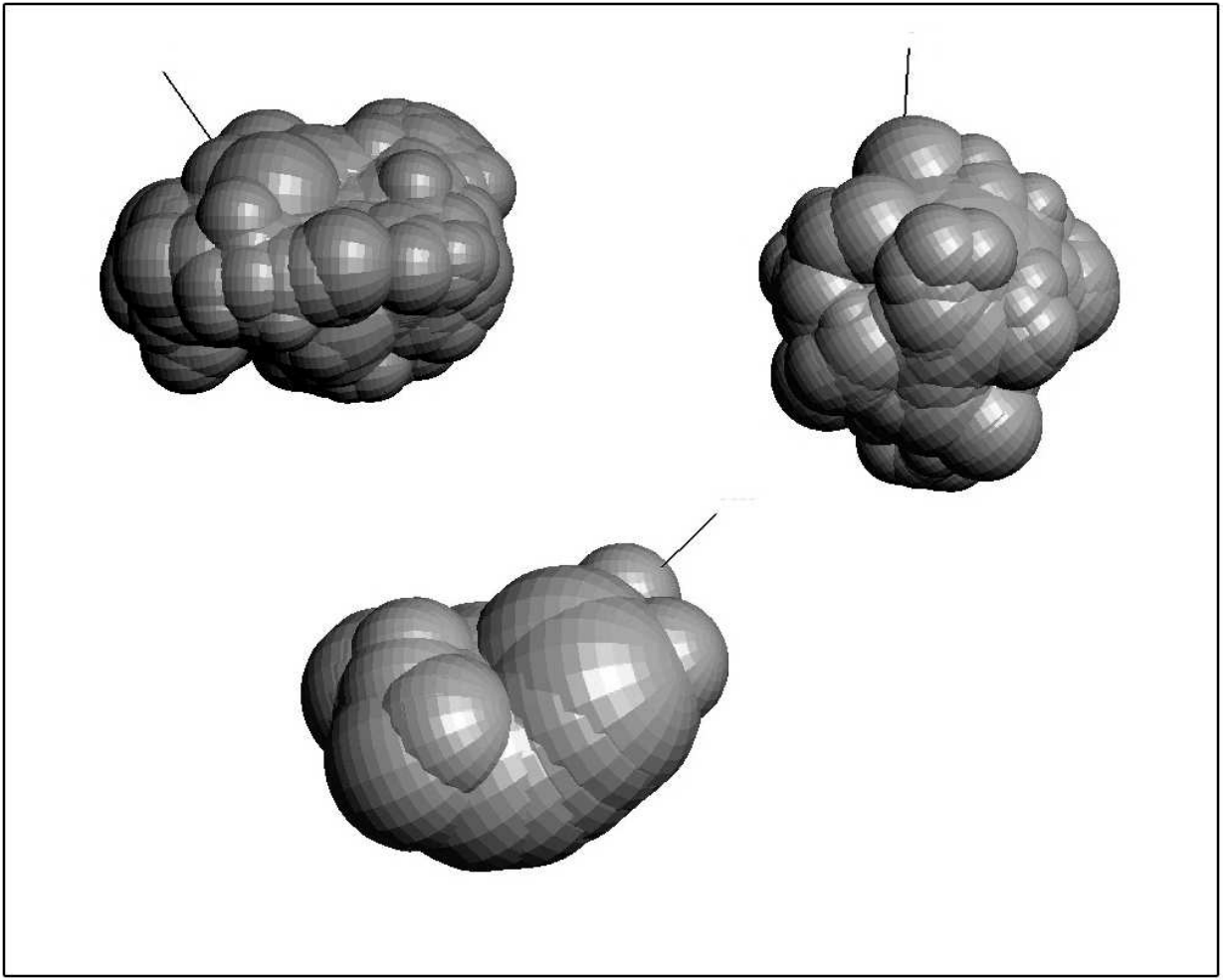}
\caption{Examples of the voids identified in the Millennium run galaxy 
catalog. The straight line on each void represents the spin axis.}
\label{fig:void}
 \end{center}
\end{figure}

We first choose only those voids which contain more than $30$ void
galaxies.  Total number of such voids is found to be $13507$. 
The specific spin angular momentum, ${\bf j}$, of each selected void is 
determined as 
\begin{equation}
\label{eqn:angnum}
{\bf j} = \frac{1}{M_{V}}\sum_{\alpha}^{N_{V}}m_{\alpha}{\bf r}_{\alpha}
\times{\bf v}_{\alpha},
\end{equation}
where $m_{\alpha}$ and $M_{V}\equiv \sum_{\alpha}^{N_{V}}m_{\alpha}$ are 
the mass of the $\alpha$-th galaxy and the total mass of all galaxies 
in each void, respectively. The position and the velocity of the 
$\alpha$-th void galaxy in the  center of mass frame are represented
by ${\bf r}_{\alpha}$ and ${\bf v}_{\alpha}$, respectively. We rescale
this specific spin angular momentum to be dimensionless as 
\begin{equation}
\label{eqn:tilj}
\tilde{j} \equiv \frac{j}{\sqrt{2}M_{V}R_{e}V},
\end{equation}
with $V^{2}\equiv GM_{V}/R_{e}$,  which is analogous to the definition of 
the halo spin parameter \citep{bul-etal01}. Figure \ref{fig:void} 
illustrates a few examples of the voids identified from the catalog with  
the directions of the void spin axes marked as short solid line on the
voids.

\begin{figure}
  \begin{center}
   \plotone{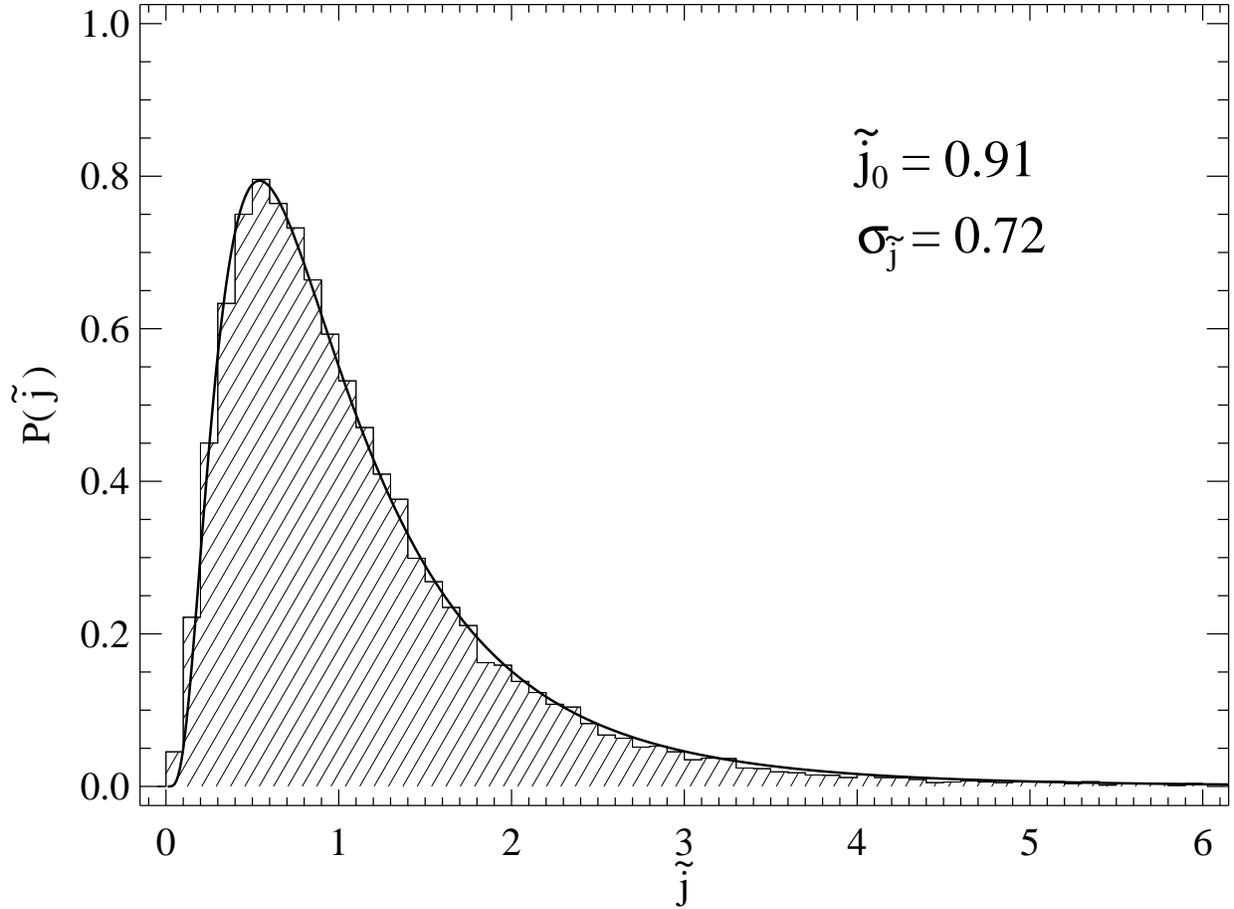}
\caption{Distribution of the rescaled spin angular momentum of voids.
The histogram represents the result from the Millennium run simulation while 
the solid line is the log-normal fit  (eq.[\ref{eqn:spindis}]).}
\label{fig:log}
 \end{center}
\end{figure}

Figure \ref{fig:log} plots the distribution of $\tilde{j}$ for the  voids 
from the catalog as histograms, which shows that $p(\tilde{j})$ has a 
log-normal shape. We fit the following log-normal formula to the histogram 
with adjusting the values of $\sigma_{\tilde{j}}$ and $\tilde{j}_{0}$:
\begin{equation}
\label{eqn:spindis}
p(\tilde{j}) = \frac{1}{\tilde{j}\sqrt{2\pi}\sigma_{\tilde{j}}}
\exp\left(-\frac{\ln^{2}(\tilde{j}/\tilde{j}_{0})}
{2\sigma^{2}_{ \tilde{j}}}\right),
\end{equation}
The best-fit values of $\sigma_{\tilde{j}}$ and $\tilde{j}_{0}$ are found to 
be $0.72$ and $0.91$, respectively. As can be seen, the value of $\tilde{j}$ 
spreads over a large range, revealing the strong shear effect on voids.

\subsection{\it Determining the Void Spin-Spin and Spin-Density Correlations}

We measure the cosine of the relative angle, $\cos\theta$, between the spin 
axes of each void pair in the catalog, and determine the average of 
$\cos^{2}\theta$ as a function of the separation distance between the 
two voids. For this, we also use only those voids which have more than $30$ 
galaxies. The two-point spin-spin correlation function (eq.[\ref{eqn:eta}]) 
is then determined simply by subtracting one third from the average, 
$\langle\cos^{2}\theta\rangle$. 

\begin{figure}
  \begin{center}
   \plotone{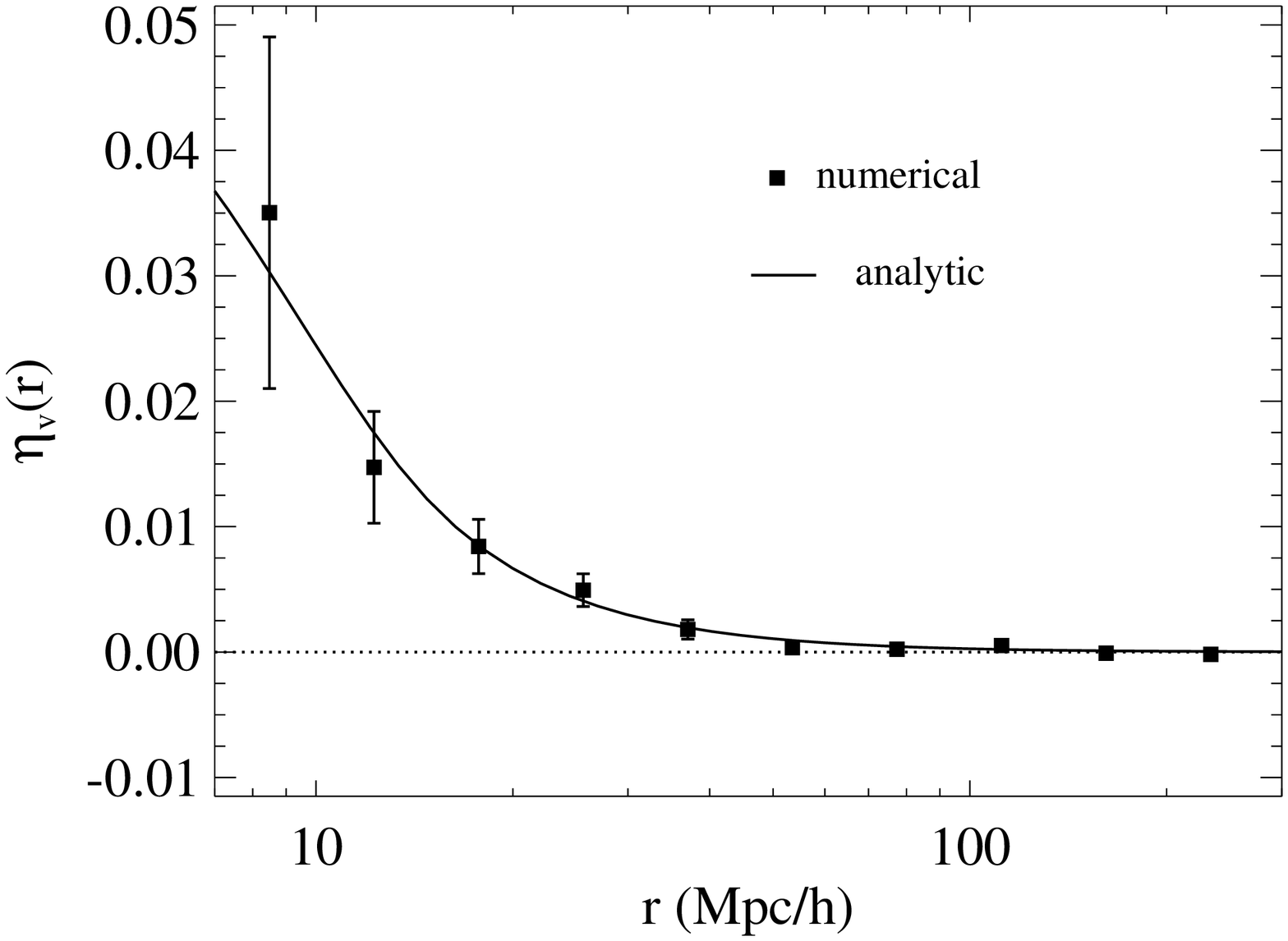}
\caption{Void spin-spin correlation function (eq.[\ref{eqn:eta}]). 
The solid squares with errors represent the result from the Millennium 
run simulation while the solid line is the analytic prediction 
(eq.[\ref{eqn:spincorrft}]). The horizontal dotted line corresponds to 
the case of no alignment.}
\label{fig:spincorr}
 \end{center}
\end{figure}

Figure \ref{fig:spincorr} plots the numerical result as solid squares 
with errors. The errors are calculated as the standard deviation of 
$\cos^{2}\theta$ for the case of no correlation. The horizontal dotted 
line corresponds to the case of no alignment. As can be seen, there is a 
clear signal of alignment up to the distance of $20h^{-1}$Mpc, which is 
almost an order of magnitude stronger compared with the case of halos.

The analytic prediction (eq.[\ref{eqn:spincorrft}]) is also plotted 
as solid line for comparison. As can be seen, the analytic prediction 
is in excellent agreement with the numerical result. It is worth emphasizing 
that the analytic prediction is not a fitting model but derived from first 
physical principles under the assumption that voids form in the initial 
regions where the tidal and the inertia tensors are maximally uncorrelated 
with each other. The good agreement between the numerical result and the 
analytical prediction supports this idea.

\begin{figure}
  \begin{center}
   \plotone{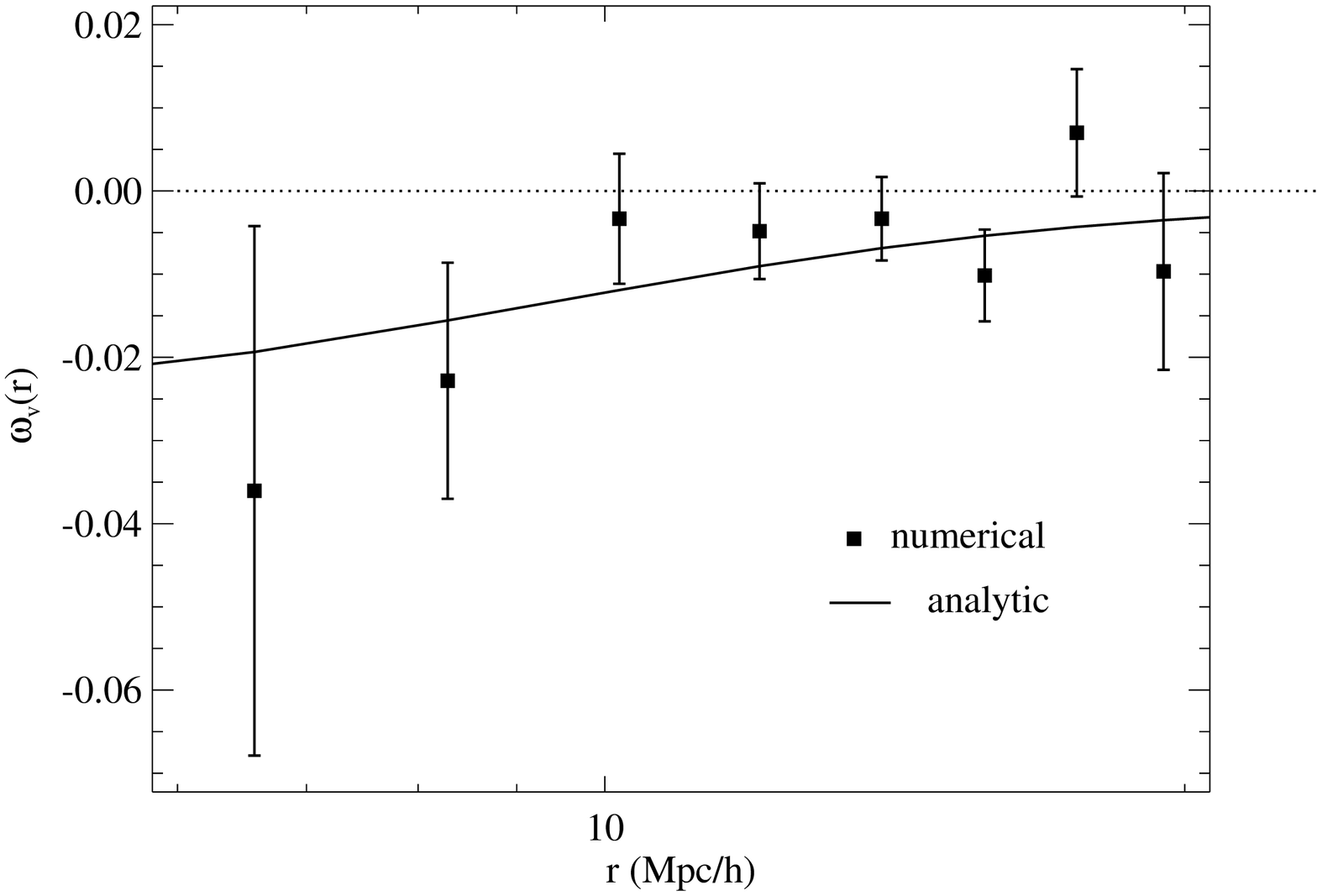}
\caption{Void spin-density correlation function (eq.[\ref{eqn:omega}]).
The solid squares with errors represent the result from the Millennium run 
simulation while the solid line is the analytic prediction 
(eq.[\ref{eqn:dencorrft}]). The horizontal dotted line corresponds to the 
case of no alignment.}
\label{fig:dencorr}
 \end{center}
\end{figure}

In a similar manner,  we also determine the void spin-direction correlation 
function (eq.[\ref{eqn:omega}]). Figure \ref{fig:dencorr} plots the numerical 
result as solid squares with errors. As can be seen, there is a clear 
signal of anti-alignment between the void spin axes and the direction 
toward the nearest neighbor voids, which is also an order of magnitude 
stronger compared with the case of halos. The analytic prediction 
(eq.[\ref{eqn:dencorrft}]) is also plotted as solid line for comparison. 
We find a good agreement between the analytical and the numerical results 
for the spin-direction correlation, too.

\section{DISCUSSION AND CONCLUSION}

In the light of the work of \citet{sha-etal06} who claimed that
the tidal shear must play a strong role in case of voids, we have proposed 
a scenario that voids are originated from the regions in the initial density 
field whose inertia tensors are completely uncorrelated with the local 
tidal tensors. Since the alignment between the principal axes of the 
inertia and tidal tensors plays a role of reducing the tidal shear effect 
according to the linear tidal torque theory, the hypothesis implies 
that voids should have very high spin angular momentum. We define the
concept of the spin angular momentum for an unbound void to quantify
the effect of the tidal field on the distribution of matter that make
up voids. 

Our analytic model based on the linear tidal torque theory have
predicted that there is a strong spatial alignment between the spin
axes of neighbor voids and a strong anti-alignment between the spin
axes of voids and the directions to the nearest voids.

Our predictions have been tested against the Millennium run galaxy catalog 
where $13507$ large voids with more than $30$ galaxies are identified.
We have measured numerically the distribution and correlation functions 
of void spins, and found that the predictions of our analytic model are in 
excellent agreement with the numerical results.

The success of our work is, however, subject to a couple of caveats. 
The first caveat comes from our implicit assumption that the matter 
voids are same as the galaxy voids. In our analytic model, we consider 
the matter voids. While in the numerical analysis we deal with the galaxy 
voids. Although the two types of voids are not necessarily same, 
they should be at least closely correlated with each other. 

The second caveat lies in the fact that there is no standard way to 
identify voids. Unlike the case of halo-finding, the use of different 
algorithms could produce different results for some properties of voids.
Especially, the shapes and volumes of voids can differ significantly by the 
underlying criterion of the void-finding algorithm. If one use an algorithm 
which regards a void as a spherical region of low density, then all voids 
found by that algorithm would be spherical. 
In spite of this ambiguity in finding voids,  we expect that our result 
on the correlation properties of the void spin angular momentum would not 
change significantly by the different choice of the void finding algorithm, 
since the correlation properties are basically determined not by the shapes 
of voids but by the galaxies in voids. 

Another issue we would like to discuss here is a possibility of
testing numerically our key assumption that a void forms in the
initial underdense region where the misalignment between the principal
axes of the inertia and the tidal tensors becomes maximum. Although
\citet{lee06} has already shown analytically that our assumption is
true,  it would be highly desirable to test it against numerical data from the
Millennium run simulation. For this, it would be necessary to find the
Lagrangian positions of all the particles that make up each void and
to determine the inertia momentum tensor of each Lagrangian proto-void
region in the principal axis frame of the initial tidal field. As
the initial density field of the Millennium run simulation is not
available to the public yet, the numerical test of our assumption is
postponed. We hope to report the test result elsewhere in the future.

Finally, we conclude that our work on void spins will provide a deeper 
insight into the origin and properties of the large scale structure in
the universe.
 
\acknowledgments

We thank the anonymous referee for helpful suggestions.
The Millennium Run simulation used in this paper was carried out by the Virgo 
Supercomputing Consortium at the Computing Center of the Max-Planck Society 
in Garching. The semi-analytic galaxy catalogue is publicly available at
http://www.mpa-garching.mpg.de/galform/agnpaper. This work is supported by 
the research grant No. R01-2005-000-10610-0 from the Basic Research Program 
of the Korea Science and Engineering Foundation.

\clearpage

\end{document}